**Identifiability and endpoint reconstruction of time-dependent Jacobi Hamiltonians from projective trajectory jets**

Alexander M. Tishin

Lomonosov Moscow State University, Physics Department, Leninskie Gory 1, 119991 Moscow, Russia

Moscow Institute of Physics and Technology, Institutskiy per. 9, 141701 Dolgoprudny, Moscow Region, Russia

Email: tishin@amtc.org

## Abstract

We study the identifiability of smooth time-dependent Jacobi matrices from phase-resolved trajectories generated by Schrödinger evolution. When the initial state is localized at a chain endpoint, the instantaneous inverse map is singular. We prove that the projective trajectory jet through order $M+1$ uniquely determines the first $M$ couplings, their logarithmic time derivatives, and the associated diagonal energy differences at preparation, and that this derivative order is necessary in the unrestricted real-Jacobi class. A complementary nonuniqueness theorem shows that observing only the first $K$ projective coordinates, even through all Taylor orders, leaves every deeper instantaneous layer undetermined for unrestricted smooth time laws. We also construct indistinguishable evolutions on a window separated from preparation, establishing ambiguity of endpoint coupling scales and rates. We identify additional observation schemes that restore local identifiability: first-order projective and reference-population data at generic times, or second-order data from $\lceil M/2 \rceil$ suitable site preparations for independent coupling rates. Statistical distinguishability bounds and numerical finite-window studies quantify the effects of noise, nuisance coefficients and acquisition design. The results delineate uniqueness and data requirements for inverse reconstruction in structured quantum dynamics, including tight-binding chains and driven bosonic ladders.

## I. Introduction

Reconstruction of a time-dependent tridiagonal operator from state evolution leads to an inverse problem with singular endpoint data. For a state initially localized at the endpoint, the first derivative probes only the nearest coupling; higher derivatives reveal coefficients farther along the chain. We determine which instantaneous Jacobi coefficients can be recovered from finite projective trajectory jets, the required derivative orders, and the limitations imposed by partial or delayed observations.

Mikhaylov and Mikhaylov [1] reconstruct finite Jacobi matrices using dynamical Dirichlet-to-Neumann data from a controlled discrete system. Here the data are physical-time derivatives of a single projective Schrödinger trajectory with smooth time-dependent Jacobi coefficients. These evolution and observation models give rise to distinct identifiability problems.

We consider continuous-time Schrödinger evolution with a fixed real-Jacobi structure and unrestricted smooth coefficient laws. Endpoint preparation makes the instantaneous state-derivative inverse map rank deficient. Theorem 1 and Corollary 1 provide a constructive reconstruction and establish the minimum derivative order required for each complete accessible layer. Theorem 2 proves that deeper instantaneous layers remain nonidentifiable even when every Taylor coefficient of a fixed coordinate prefix is known, provided the time dependence is unrestricted and smooth. Equality of the observed jets need not imply equality of trajectories on an open interval.

The uniqueness results require the stated structural prior, phase-sensitive access and differentiability assumptions. Proposition 2 addresses an unobserved interval following preparation; separate numerical and information-theoretic analyses study finite-window conditioning and acquisition cost under specified statistical models. The statistical results do not establish uniform stability under noisy observations.

Projective geometry describes state distinguishability through the Fubini–Study metric [2], relates trajectory speed to energy uncertainty [3], and organizes non-adiabatic response through the quantum geometric tensor [4]. Quantum Frenet–Serret constructions describe higher curvature and torsion [5, 6]. These quantities alone do not identify a general generator from its path.

Lanczos recursion maps a known Hermitian generator to a tridiagonal chain [7, 8]. Operator Krylov geometry has coherent-state constructions [9], while state spread complexity uses the state Krylov chain [10]. Takahashi and del Campo extend chain constructions to $H(t) - i\hbar\,\partial_t$, with basis transport [11], and use operator Krylov space for counterdiabatic control [12]. A Toda–Arnoldi construction instead concerns a fixed Hamiltonian and a holomorphically deformed seed [13]. Classical inverse Jacobi theory includes dynamical boundary-response and boundary-control reconstruction [14]. Our inverse map separates instantaneous coefficient values and rates from a physical-time endpoint jet.

Hamiltonian identification uses structural priors [15], measurement time traces [16, 17], quantum resources [18], sparse or local models [19 - 21], and continuous weak monitoring of time-dependent generators [22]. Restricted-access methods recover static spin-chain parameters from one end [23 -25], and boundary tomography extends to Bose–Hubbard graphs [26]. In a static chain, moment inversion recovers Lanczos coefficients [27, 28]. General prescribed-path constructions also exhibit nonuniqueness without a structural prior [29].

For time-dependent systems, neural-network approaches use multiple known initial states and observable traces [30]. Short-time learning studies measurement design and bias/noise competition [31]. Prepare–evolve–measure schemes learn local generators and noise under finite-dimensional time-law priors [32], and continuous monitoring can exploit sparse interactions [33]. Constrained parent-Hamiltonian reconstruction expresses the generator as an instantaneous state-derivative inverse problem [34]. For a fixed real-Jacobi class, we solve the singular endpoint problem explicitly and determine the derivative order required to recover a complete layer.

For endpoint initialization, the phase-resolved jet through order $M + 1$ recovers $M$ complete layers. Classical generalized Frenet and moment/Lanczos constructions supply the frame and static dictionary [35-37]. The Supplement gives the derivative-QR representation and its distinction from state spread bases. A pair-driven oscillator provides a physical realization, with vacuum evolution confined to the even-Fock ladder [38- 41].

The assumed access consists of coherences between the reference and every observed level, together with the reference population. It is sufficient for the unrestricted-drive theorem and differs from end-only tomography. Proposition 2 gives a complementary obstruction: continuous full-state observation after an unobserved preparation gap leaves endpoint scales ambiguous in the unrestricted smooth class. Quantitative time-law restrictions are therefore part of finite-window estimation. We compare access through additional site preparations and later observation windows, including the effects of finite-window bias, nuisance coefficients and acquisition errors. Numerical methods and the relation to established constructions are given in the Supplemental Material.

## II. Endpoint inverse reconstruction

Let $\{e_n\}_{n=0}^{N-1}$ be a fixed orthonormal ladder. In an open interval containing a reconstruction time $t_0$, assume a smooth real Jacobi Hamiltonian

$$H(t) = \sum_{n=0}^{N-1} E_n(t) e_n e_n^\dagger + \sum_{n=0}^{N-2} g_n(t)\left(e_{n+1} e_n^\dagger + e_n e_{n+1}^\dagger\right), \tag{1}$$

with $g_n(t) > 0$ over the depth to be reconstructed. The theorem is stated in finite dimension. For a semi-infinite ladder the same finite-depth identities hold provided a self-adjoint realization and a strong Schrödinger evolution are specified, the endpoint solution is differentiable through order $M + 2$, and the differentiated Schrödinger equation is valid through the orders used. The required mixed products of $H$ and its time derivatives must be well defined in their operator domains. Order $M + 2$ supplies the displayed big-$O$ remainders; the coefficient identities require strong derivatives through $M + 1$. These are assumptions of the extension. Smooth matrix entries alone are insufficient to guarantee them. Nearest-neighbour support makes the endpoint identities depend on finitely many ladder coefficients. Existence and uniqueness of unbounded evolution require the stated hypotheses. The reconstruction time is the time of endpoint preparation. Applying this result at another physical time requires a corresponding endpoint preparation; the evolved state at a generic later time need not satisfy this condition. The state obeys

$$i\hbar\dot{\psi}(t) = H(t)\psi(t), \psi(t_0) = e_0. \tag{2}$$

A scalar addition $s(t)I$ changes only the global phase and is therefore invisible in projective dynamics. Write $\psi = \sum_n c_n e_n$ and use the local chart

$$z_n(t) = \frac{c_n(t)}{c_0(t)}, z_n(t_0) = 0 (n \geq 1). \tag{3}$$

For $n \geq 1$ define

$$A_n = z_n^{(n)}(t_0), R_n = \frac{z_n^{(n+1)}(t_0)}{z_n^{(n)}(t_0)}, \quad A_0 = 1, R_0 = 0. \tag{4}$$

Theorem 1 Endpoint projective jet reconstruction. Let $1 \leq M \leq N-1$. For the real Jacobi chain (1) initialized by (2), with $g_0, \dots, g_{M-1}$ positive in a neighbourhood of $t_0$, the jets of $z_1, \dots, z_M$ through order $M+1$ determine the first $M$ instantaneous ladder layers. For $1 \leq n \leq M$,

$$A_n = \left(-\frac{i}{\hbar}\right)^n \prod_{j=0}^{n-1} g_j\,(t_0), \tag{5}$$

and

$$R_n - R_{n-1} = n\frac{\dot{g}_{n-1}(t_0)}{g_{n-1}(t_0)} - \frac{i}{\hbar}[E_n(t_0) - E_0(t_0)]. \tag{6}$$

Consequently,

$$g_{n-1}(t_0) = \hbar\left|\frac{A_n}{A_{n-1}}\right|, \tag{7}$$

$$E_n(t_0) - E_0(t_0) = -\hbar \mathrm{Im}(R_n - R_{n-1}), \tag{8}$$

and

$$\frac{\dot{g}_{n-1}(t_0)}{g_{n-1}(t_0)} = \frac{1}{n}\mathrm{Re}(R_n - R_{n-1}). \tag{9}$$

Within the reconstructed layers, the scalar shift $E_0(t)I$ is the only unavoidable diagonal ambiguity. If the next ladder level exists and the additional coordinate $z_{M+1}$ is observed through order $M+1$, its jet also supplies the onset coefficient $A_{M+1}$ and therefore fixes $|g_M(t_0)|$, including zero, through (7); Theorem 1 counts complete layers, each consisting of a coupling, its logarithmic rate and the energy offset reached through it. The proof, given in Sec. VIII, uses the unique minimal path $0 \to 1 \to \cdots \to n$: level $n$ first appears at order $t^n$, and the next local order separates the logarithmic coupling rate from the diagonal energy difference.

Corollary 1 (sharpness of the reconstruction order). In the unrestricted real Jacobi class of Theorem 1, order $M+1$ is necessary in general to identify $E_M(t_0) - E_0(t_0)$. Two Hamiltonians differing only by a nonzero constant diagonal shift $\delta e_M e_M^\dagger$ have identical full projective jets through order $M$. Writing $\tau = t - t_0$, their first difference is

$$\tilde{z}_M(t) - z_M(t) = \frac{(-i)^{M+1}\delta}{(M+1)!\,\hbar^{M+1}}\left(\prod_{j=0}^{M-1} g_j\,(t_0)\right)\tau^{M+1} \\ +O(\tau^{M+2}). \tag{10}$$

The couplings at depth $M$ already enter at order $M$, whereas the deepest diagonal energy and coupling rate require the next order. Additional relations between Hamiltonian parameters may reduce the information needed in a more restricted model. The proof of order necessity is given in Sec. VIII B.

The assumption that the chain is real in a neighbourhood of $t_0$ is substantive. For a Hermitian complex extension, define the lower hop $g_{n-1} = \langle e_n|H|e_{n-1}\rangle = \rho_{n-1}e^{i\phi_{n-1}}$. Its contribution to the operator is $g_{n-1}|e_n\rangle\langle e_{n-1}| + g_{n-1}^{*}|e_{n-1}\rangle\langle e_n|$. With this orientation the same recursion gives

$$R_n - R_{n-1} = n\frac{\dot{\rho}_{n-1}}{\rho_{n-1}} + in\dot{\phi}_{n-1} - \frac{i}{\hbar}(E_n - E_0). \tag{11}$$

Thus $|g_{n-1}|$ and its logarithmic amplitude derivative remain accessible, but the imaginary part determines the energy only after the local coupling-phase rate is specified. Equivalently, a time-dependent chain gauge can remove the coupling phases at the price of shifting diagonal energies. We therefore keep Theorem 1 in the fixed real-chain gauge and treat (11) as the gauge-covariant extension.

Theorem 2 (all-order limits of partial endpoint data). Fix $1 \le K \le N-1$ and a smooth real-Jacobi germ with positive endpoint couplings. The full physical-time Taylor jets of $z_1, \dots, z_K$ determine the first $K$ complete instantaneous layers. If $K < N-1$, the jets of every deeper coupling $g_j$, $j \ge K$, and diagonal offset $E_n - E_0$, $n > K$, can instead be prescribed arbitrarily, with positive coupling values at preparation, while preserving all observed jets. The higher derivatives of the first $K$ layers adjust to these choices. Thus deeper instantaneous layers are not identifiable from these partial jets in the unrestricted $C^\infty$ class.

Proof. Fix $E_0 = 0$ and write $[t^m]z_n$ for Taylor coefficients. At order $m \ge n+1$, the fresh coefficient of $g_{n-1}$ at order $m-n$ and the fresh coefficient of $E_n$ at order $m-n-1$ enter $[t^m]z_n$ with factors

$$\begin{aligned} a_{nm} &= \frac{(-i)^n \prod_{j<n-1} g_j\,(0)}{\hbar^n (n-1)!\,m}, \\ b_{nm} &= \frac{(-i)^{n+1} \prod_{j<n} g_j\,(0)}{\hbar^{n+1} n!\,m}. \end{aligned} \tag{12}$$

They are nonzero and in phase quadrature. Neither fresh coefficient changes another observed coordinate at order $m$ or a previously fixed order. After prescribing the hidden jets, solve these two real equations for each observed $n$, order by order. Leading-order coefficients and Theorem 1 fix the first $K$ layers. Borel's extension theorem [42] realizes the resulting real coefficient sequences by smooth functions; continuity preserves positivity near preparation. The finite-dimensional Schrödinger recurrence then gives the prescribed observed jets. The Supplement gives the path argument and the extension details. The result asserts equality of Taylor jets. Distinct smooth trajectories may have those jets. In polynomial subclasses, additional time-law relations can make one coordinate sufficient for local identification.

### A. Relation to constrained inverse problems

The constrained state-derivative inverse problem of Rattacaso et al. [34] determines the visible generator component and leaves operators stabilizing the state in its instantaneous kernel. Theorem 1 specializes that framework to the singular endpoint of a fixed real Jacobi class. Choose allowed operators $D_j = |e_j\rangle\langle e_j|$ and $B_j = |e_j\rangle\langle e_{j+1}| + |e_{j+1}\rangle\langle e_j|$. The covariance matrix $V_{ab} = \langle L_a L_b + L_b L_a\rangle - 2\langle L_a\rangle\langle L_b\rangle$ at $\psi(t_0) = e_0$ has only $V(B_0, B_0) = 2$ nonzero, hence rank one. The first derivative fixes $g_0$, while other layers are initially invisible. This says nothing about the information in the whole later trajectory. In a three-level chain with $E_0 = 0$ and $\hbar = 1$, write $g_0 = g, g_1 = h, E_1 = E, E_2 = F$. The identities give $A_1 = -ig, A_2 = -gh, R_1 = \dot{g}/g - iE$ and $R_2 - R_1 = 2\dot{h}/h - iF$. Order two therefore determines $g, h, E, \dot{g}/g$, while order three also determines $F, \dot{h}/h$. The static limit has vanishing real parts and retains the familiar inverse-chain structure [35, 36, 37]. The specific result is the closed higher-order triangular recursion, its scalar-energy gauge and its necessary order, not the general observation that more trajectory data constrain a parent Hamiltonian. Finite samples do not determine classical derivatives exactly without approximation control.

The following projective constraint gives the algebraic connection to that framework. Writing $P = |\psi\rangle\langle\psi|$, its projective parent constraint is

$$(I - P)\big(H\psi - i\hbar\dot{\psi}\big) = 0. \tag{13}$$

In a chart with nonzero reference amplitude, eliminating the scalar component gives the ratio equation (27). Its differentiated endpoint constraints are solved by matching the coefficients of $t^{n-1}$ and $t^n$, namely (28) and (29) (Sec. VIII). Since $z_n = O(t^n)$, an energy derivative multiplies it first at right-hand-side order $t^{n+1}$, affecting $z_n$ only at order $t^{n+2}$; in contrast, $\dot{g}_{n-1}t$ multiplies $z_{n-1} = O(t^{n-1})$ and enters the next coefficient used here. The resulting ratios have the closed form

$$R_n = \sum_{j=0}^{n-1}(j+1)\frac{\dot{g}_j}{g_j} - \frac{i}{\hbar}\sum_{k=1}^{n}(E_k - E_0)\,. \tag{14}$$

Subtracting $R_{n-1}$ cancels all earlier weighted rates and energies, leaving $n\dot{g}_{n-1}/g_{n-1}$ and $E_n - E_0$. The explicit differentiated solution gives the necessary layerwise information order within constrained parent-Hamiltonian reconstruction.

### III. Local distinguishability and accessible depth

Proposition 1 (local statistical distinguishability). For a positive ladder depth $n$, with $1 \le n \le L \le N - 1$, consider two finite real Jacobi chains under Theorem 1 that differ only by the constant diagonal shift $\delta e_n e_n^\dagger$, with fixed nonzero $\delta$. Observe any fixed set of coordinates $z_1, \dots, z_L$, $L \ge n$, at $t_\ell = t_0 + Tx_\ell$, where $K$ fixed sampling fractions satisfy $0 < x_\ell \le 1$. Suppose each measured real and imaginary component has independent Gaussian noise of known standard deviation $\sigma$. Define

$$\kappa_n = \frac{\prod_{j=0}^{n-1} g_j\,(t_0)}{(n+1)!\,\hbar^{n+1}}, w_n = \sum_{\ell=1}^{K} x_\ell^{2n+2}. \tag{15}$$

Let $\mu$ and $\tilde{\mu}$ be the real vectors of all noise-free coordinate components and define $\mathcal{D} = \parallel \tilde{\mu} - \mu \parallel_2/\sigma$. As $T \to 0$ with the Hamiltonians, sample fractions and measured depth fixed,

$$\mathcal{D} = \frac{|\delta|\kappa_n T^{n+1}\sqrt{w_n}}{\sigma}[1 + O(T)], P_{err}^* = \Phi(-\mathcal{D}/2). \tag{16}$$

At leading order only one quadrature of $z_n$ carries the shifted-energy signal; higher observed levels contribute at higher powers of $T$. Here $\Phi$ is the standard normal cumulative distribution function, and $P_{err}^*$ is the exact optimal error probability for testing the two fixed Gaussian models with equal prior probabilities. The local Fisher information for the diagonal shift, with every nuisance parameter held fixed, is $\mathcal{I}_n = \kappa_n^2 T^{2n+2} w_n/\sigma^2[1 + O(T)]$. Consequently, the standard deviation $sd(\hat{\delta})$ of any locally unbiased regular estimator $\hat{\delta}$ of $\delta$ obeys the optimistic Cramér–Rao bound

$$sd(\hat{\delta}) \geq \mathcal{I}_n^{-1/2} = \frac{\sigma}{\kappa_n T^{n+1}\sqrt{w_n}}[1 + O(T)]. \tag{17}$$

Unknown nuisance parameters cannot improve on the information obtained when they are fixed. The binary bound applies without an unbiasedness assumption, whereas (17) has the usual regularity and local-unbiasedness conditions. A weak intermediate coupling reduces $\kappa_n$ and therefore limits accessible depth even before it vanishes. These are observation-model-dependent local bounds, not a universal experimental noise threshold. The short-time expansion holds for fixed time laws; a small $T\max_j\{|g_j|/\hbar, |E_j - E_0|/\hbar, |\dot{g}_j/g_j|\}$ describes instantaneous scales but does not control arbitrarily large higher derivatives. The remainder constants depend on those derivatives. Correlated errors require the corresponding covariance-weighted distance and Fisher information.

## IV. Observation gaps and model restrictions

Proposition 2 (endpoint ambiguity from an observation gap). Let $H(t)$ be a $C^\infty$ finite-dimensional real Jacobi Hamiltonian with positive couplings, prepared at $t = 0$. Suppose the observed window is $[\tau, T]$, with $\tau > 0$. For every $a > 0$ there is a smooth real Jacobi Hamiltonian $\tilde{H}$ with positive couplings that gives the same entire state trajectory on that window, while

$$\tilde{g}_j(0) = a g_j(0), \tilde{E}_n(0) - \tilde{E}_0(0) = a[E_n(0) - E_0(0)]. \tag{18}$$

Thus any observation channel depending only on the freely evolved state on $[\tau, T]$, including a continuous full-state record, has the same law for distinct endpoint coefficients. Additional control interventions or direct drive measurements define a different experiment.

Proof. Choose a smooth increasing map $f$ with $f(0) = 0$, $f'(0) = a$ and $f(t) = t$ for $t \geq \tau$. Such a map exists by choosing a positive smooth derivative with endpoint value $a$, equal to one near $\tau$, and integral $\tau$ on $[0, \tau]$; its near-zero support can be shortened for arbitrarily large $a$. Direct differentiation gives

$$\tilde{H}(t) = f'(t)H(f(t)), \tilde{\psi}(t) = \psi(f(t)). \tag{19}$$

The two Hamiltonians evolve under the same physical clock $t$. The second equation solves the Schrödinger equation for the first, preserves endpoint preparation, and equals $\psi(t)$ throughout the observed window. Positive $f'$ preserves the real-chain prior. Evaluating at zero gives (18). This construction changes physical couplings and energy differences. The exact endpoint jet changes, consistently with Theorem 1. The endpoint ratios $g_j/g_0$ and $(E_n - E_0)/g_0$ are invariant under this witness. Its logarithmic rate transforms as $\tilde{\chi}_j(0) = a\chi_j(0) + f''(0)/a$. Thus the witness establishes scale and rate ambiguity, not arbitrary independent ambiguity of dimensionless ratios.

The construction uses compactly supported smooth changes. It does not stay within a fixed polynomial or analytic time-law class. In a finite-dimensional analytic real-Jacobi class, continuous noiseless trajectory data can in principle be continued to preparation, although this continuation may be ill-conditioned. A quasi-analytic extension requires the corresponding uniqueness and evolution-closure hypotheses. A prescribed nonzero constant diagonal coefficient, such as a fixed Kerr term, is generally changed by the rescaling; the witness is not a coupling-only pump change in that subclass.

Corollary 2 (a derivative-bounded error floor). Let $H_*$ be a constant positive-coupling real Jacobi matrix, $B = \min_s \| H_* - sI \| > 0$, and $\theta_* = E_n - E_0 \neq 0$. In the class of $C^\infty$ real-Jacobi drives with positive couplings and $\| \dot{H}(t) \| \leq L$ on $[0, T]$, with $L > 0$, the state-only experiment on $[\tau, T]$ has minimax expected absolute error at least $\varepsilon_{\mathrm{w}}|\theta_*|/2$ and squared error at least $\varepsilon_{\mathrm{w}}^2\theta_*^2/4$, where

$$\varepsilon_{\mathrm{w}} = \min\{1/2, L\tau/(K_2 B)\}, K_2 \simeq 4.6461. \tag{20}$$

The Supplement proves the constant using the extrema of a smooth bump. For the control below, $B \simeq 2.334$, $L \simeq 0.3065$ and $\tau = 0.024$ give an absolute-error floor $2.37 \times 10^{-4}$, below its observed estimator errors. This lower bound concerns a constructed subclass. Exact identifiable data at $\tau$ give the conditional upper estimate discussed next.

### A. Generic-time data and endpoint data

Proposition 3 (generic-time reconstruction under specified amplitude access). At a given time suppose every amplitude $c_n$ is nonzero and $\mathrm{Im}(c_{n+1}/c_n) \neq 0$ for every adjacent pair. The full state and its physical-time derivative determine a real Jacobi Hamiltonian in the fixed basis uniquely when a Schrödinger lift is specified, and up to a scalar when only the projective tangent is specified.

The prefix version needs nonzero amplitudes and adjacent phase conditions only in the observed prefix. For $1 \leq L < N$, amplitudes $c_0, \ldots, c_L$ and their derivatives in a Schrödinger lift determine $g_0, \ldots, g_{L-1}$ and $E_0, \ldots, E_{L-1}$, using only rows $0, \ldots, L-1$. With projective data, the same conclusion up to a scalar uses the ratios $z_1, \ldots, z_L$ together with the reference population $p_0$ and their first derivatives.

Proof. If a real Jacobi matrix $K$ annihilates $c$, its first row gives $d_0 + b_0 c_1/c_0 = 0$. The imaginary part gives $b_0 = 0$, then $d_0 = 0$. The same argument applied successively to the remaining rows gives all coefficients zero, including the last diagonal. For a projectively invisible matrix, $Kc = \lambda c$ with real $\lambda$; apply the preceding argument to $K - \lambda I$.

A constructive recursion sets $y_n = i\hbar \dot{c}_n / c_n$ and $g_{-1} = 0$. For $n < N-1$,

$$\begin{aligned} g_n &= \frac{\mathrm{Im}(y_n - g_{n-1} c_{n-1}/c_n)}{\mathrm{Im}(c_{n+1}/c_n)}, \\ E_n &= \mathrm{Re}(y_n - g_{n-1} c_{n-1}/c_n - g_n c_{n+1}/c_n). \end{aligned} \tag{21}$$

The nonexistent term at $n = 0$ is zero; the last row gives $E_{N-1} = \mathrm{Re}(y_{N-1} - g_{N-2} c_{N-2}/c_{N-1})$. A projective lift first fixes a scalar gauge. Small amplitudes or adjacent imaginary ratios can make this recursion arbitrarily ill-conditioned.

In particular,

$$g_0 = \frac{\hbar \dot{p}_0}{2 p_0 \mathrm{Im} z_1}. \tag{22}$$

Near preparation, $\dot{p}_0 = -2 g_0^2 t/\hbar^2 + O(t^2)$ is small. Choose the local lift $c_0 = \sqrt{p_0}$ and $c_n = \sqrt{p_0} z_n$ and apply the recursion. Without this population derivative, the first $L-1$ projective rows supply $2(L-1)$ real equations for $2L-1$ real coupling/offset unknowns. The unobserved-tail row cannot close that one-parameter freedom. The full-projector case supplies it through normalization. This is a structured specialization of the state-derivative inverse map of Rattacaso et al. [34].

For exact identifiable data at $\tau$ and $\| \dot{H} \| \leq L$, using $E_n(\tau) - E_0(\tau)$ as an endpoint estimate gives error at most $2L\tau$. This matches the lower-bound dependence on $L\tau$ in the uncapped witness subclass with fixed $|\theta_*|/B$. It requires the access and nondegeneracy conditions at $\tau$. Uniform noisy stability over $\mathcal{H}_L$ requires additional hypotheses.

At small positive times after endpoint preparation, $c_{n+1}/c_n = -ig_n t/[(n+1)\hbar] + O(t^2)$, so the adjacent phase condition holds locally. Small amplitudes still impair conditioning: a five-level static control gives a condition-number slope about $-4.00$ over the tested small-time interval for the unscaled projective state-derivative map. This links the singular endpoint to depth-dependent loss of sensitivity. The Supplement specifies this example. Its exponent depends on the chosen data norm.

## V. Numerical estimation and acquisition

### A. General Jacobi control and finite-window estimation

An independent six-level real chain has endpoint energies $E = (0.12, 0.37, 1.91, 0.82, 3.44, 1.27)$, couplings $g = (0.73, 1.11, 0.58, 0.94, 0.67)$, and quadratic coefficient laws. The numerical units are $t = \Omega t_{\mathrm{lab}}$ and $H = H_{\mathrm{lab}}/(\hbar\Omega)$ for an arbitrary reference frequency $\Omega$, with $\hbar = 1$ in the calculations. Their derivatives are $\dot{E} = (0.07, -0.13, 0.05, 0.19, -0.11, 0.03)$, $\ddot{E} = (0.02, -0.01, 0.04, -0.03, 0.01, 0.05)$, $\dot{g} = (-0.19, 0.08, 0.14, -0.05, 0.09)$, and $\ddot{g} = (0.06, -0.02, 0.03, 0.04, -0.01)$. Thus each coefficient is its endpoint value plus its first derivative times $t$ and half its second derivative times $t^2$. Exact series division and Taylor propagation through sixth order recover all five accessible complete layers with maximum absolute error $8.9 \times 10^{-16}$ in double precision. This checks the generic derivative identities without the bosonic polynomial prior.

For finite-window recovery, independently integrated trajectories are sampled at 61 uniformly spaced times in $[T/10, T]$. The estimator is supplied neither a bosonic relation nor the exact quadratic coefficient law. Complex least squares fits $z_n = \sum_{k=n}^{n+p} a_{nk}\, t^k$ for $n = 1,2,3$, with $p = 2$ or 3 and rescaled time. The known endpoint onset is enforced. The fitted coefficients give $A_n = n!\, a_{nn}$, $R_n = (n+1)a_{n,n+1}/a_{nn}$ in (7)–(9). No estimates are clipped or removed; Supplemental Note 4 discusses the associated denominator tails. Seven window endpoints, three additive Gaussian noise levels and 200 trials per degree give 8,400 reconstructions. Figure 1 summarizes the degree-$n+2$ errors. DOP853 integration uses tight tolerances; tightening them changes the tested coordinates by less than $7 \times 10^{-15}$. The detailed estimator and convergence settings are provided in the Supplemental Material.

The unregularized fitted-coefficient ratios have generally divergent second moments under Gaussian coordinate noise. We report median absolute errors and interquartile bands; the Supplement distinguishes these from local delta-method scales and gives the independent integration and trial procedures.

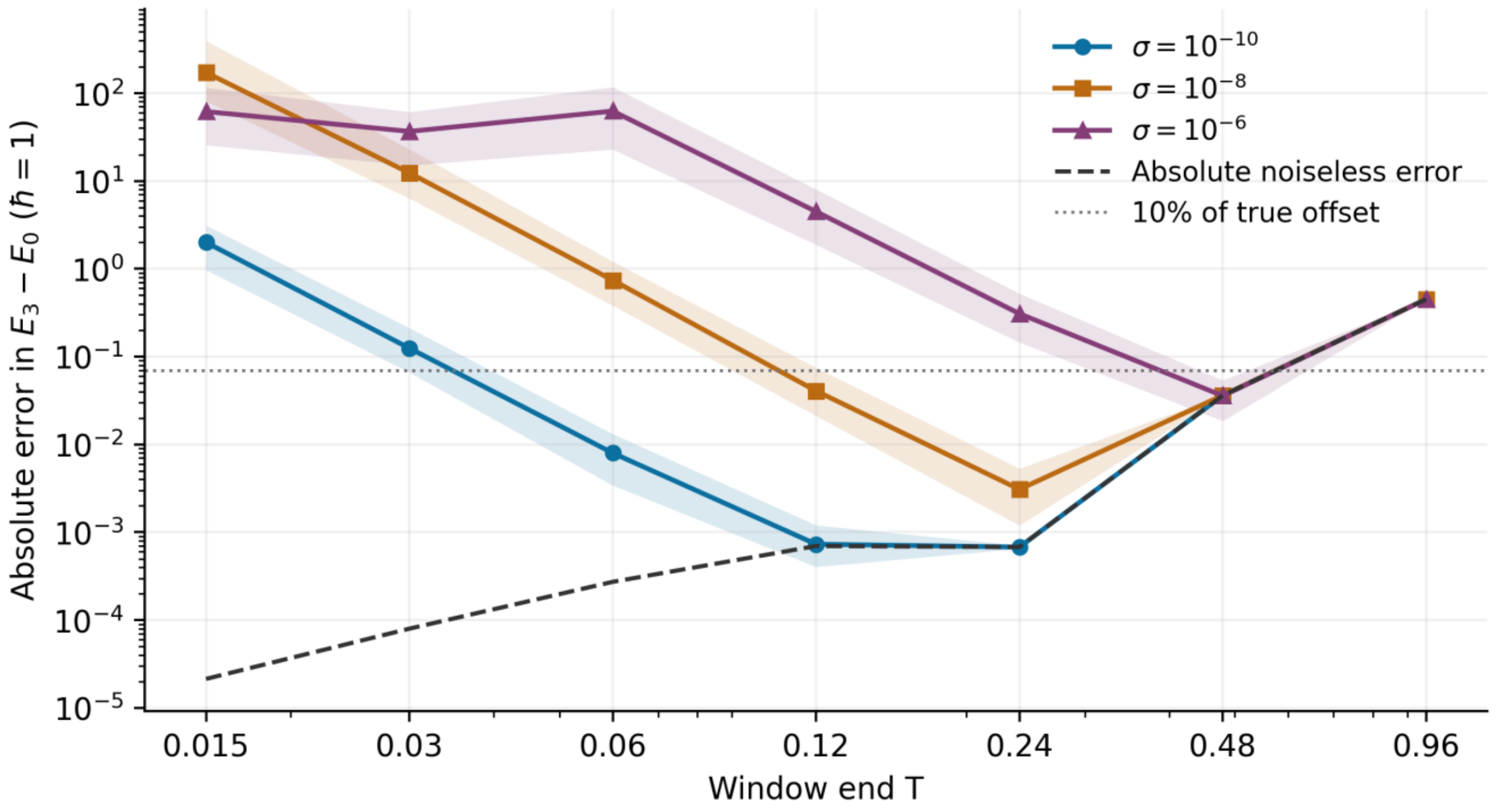

Figure 1. Noise and finite-window bias in the general Jacobi control. Median absolute third-energy error and interquartile bands from 200 realizations at each window/noise setting. The estimator retains degrees $n$ through $n + 2$. The dashed curve is the absolute error on noiseless data. The dotted line marks the selected 10% accuracy target for the true offset 0.70 in units $\hbar = 1$. The horizontal axis is logarithmic; successive tested windows double in duration. Each window contains 61 points in $[T/10, T]$. The plotted noiseless error is absolute; the signed bias changes between $T = 0.21$ and 0.24. The small bias near $T = 0.24$ partly reflects cancellation of the two constituent ratio biases.

At $\sigma = 10^{-8}$, degree $n + 2$ median absolute third-energy errors at $T = 0.12, 0.24, 0.48$ are 0.0406, 0.00308 and 0.0362, respectively; the true offset is 0.70. Percentile 95% bootstrap intervals are $[0.0347, 0.0500]$, $[0.00244, 0.00355]$ and $[0.03614, 0.03626]$, from 20,000 resamples of the 200 stored trials per group. These quantify Monte Carlo uncertainty of sample medians, not a Hamiltonian confidence interval. The intermediate tested window is better than its two neighbours in this example, not a universal optimal window. Degree $n + 3$ increases the median at $T = 0.24$ to 0.0169, with interval $[0.0136, 0.0200]$.

The error curve mixes statistical amplification with deterministic approximation bias. At $T = 0.24$, the noiseless biases of $-\mathrm{Im}R_2$ and $-\mathrm{Im}R_3$ are 0.01782 and 0.01850, leaving a third-energy bias of only 0.000679 after subtraction. A deterministic sweep gives signed third-energy biases $-0.0001763$ at $T = 0.21$ and 0.0006794 at 0.24. The minimum therefore partly reflects cancellation; it is not uniformly accurate recovery of both ratio coefficients. Bootstrap uncertainty does not measure this truncation bias.

The complete-layer output also includes its coupling and logarithmic rate, not only the favourable energy difference. Table I summarizes all three quantities from the same Gaussian trials. Their true values are $g_2 = 0.58$, $E_3 - E_0 = 0.70$, and $\chi_2 \equiv \dot{g}_2/g_2 = 0.14/0.58 \simeq 0.2414$. Relative errors use the absolute value of each corresponding truth. At $T = 0.24$, all 200 trials meet the joint 10% criterion under the simulated quadratic truth; this supplies no unrestricted-smooth-class guarantee; at the wider window the rate misses that criterion in every trial. These are observed finite Monte Carlo counts for this chain, degree and noise level, not guaranteed success probabilities or evidence of efficient estimation over arbitrary coefficient laws.

Table I. Third complete layer in the finite-window Jacobi control: median absolute relative errors (percent), with joint success counts. The estimator uses degree $n + 2$, 61 times in $[T/10, T]$, and $\sigma = 10^{-8}$. A success requires each of the three errors to be at most 10%; all 200 trials per window are included. The $T = 0.24$ point is near a deterministic energy-bias sign change.

| $T$ | $g_2$ | $E_3 - E_0$ | $\chi_2$ | Joint successes |
|---|---|---|---|---|
| 0.12 | 0.0528 | 5.807 | 6.428 | 101/200 |
| 0.24 | 0.0349 | 0.439 | 3.768 | 200/200 |
| 0.48 | 0.2794 | 5.173 | 14.527 | 0/200 |

The favourable $\sigma = 10^{-8}$ Gaussian setting corresponds near preparation to $S \simeq 2.5 \times 10^{15}$ per setting under the equal-allocation protocol, or $427S \simeq 1.07 \times 10^{18}$ preparations (about 33,800 years at 1 MHz). This is a protocol conversion; away from preparation the actual populations determine the variance.

**B. Dependence on the nuisance model**

The near equality between efficient and fixed-nuisance information depends strongly on which time-law coefficients are known. For the same quadratic control and observations, the target is a constant shift of $E_3$, with $E_0$ and its derivatives fixed. We consider nested parameter sets: S1 contains $E_1, E_2, E_3, g_0, g_1, g_2$ at the endpoint; S2 adds their first derivatives; S3 contains $E_1, \ldots, E_4, g_0, \ldots, g_3$, including each first and second derivative. All unlisted coefficients, including $E_5, g_4$ and their derivatives, are fixed. All parameters other than the target are nuisance. S3 is a finite quadratic class, not an unrestricted smooth-Hamiltonian model.

The column-normalized Jacobians have full ranks 6, 12 and 24. The S3 smallest singular values are about $9.01 \times 10^{-10}$, $7.84 \times 10^{-9}$ and $7.49 \times 10^{-8}$ for the three windows. High-precision Taylor sensitivity projections give the values in Table II; independent ODE convergence alone can miss such weak residuals. Q3 adds second derivatives of the first three layers, +d0 adds $E_4, g_3$, and +d1 adds their first derivatives. S3 also includes their second derivatives. All nuisance assumptions are finite-dimensional time-law priors.

Table II. Standard-deviation inflation relative to fixed-nuisance information. S2 has values and first derivatives of the first three layers; Q3 adds their second derivatives. Columns +d0 and +d1 add $E_4, g_3$ and then their first derivatives; S3 also adds their second derivatives.

| $T$ | S2 | Q3 | +d0 | +d1 | S3 |
|---|---|---|---|---|---|
| 0.12 | 10.0 | 55.3 | 56.5 | 224 | $1.70 \times 10^3$ |
| 0.24 | 10.1 | 55.6 | 57.9 | 228 | $1.32 \times 10^3$ |
| 0.48 | 10.1 | 56.2 | 61.6 | 238 | 870 |

At $T = 0.24$ and $\sigma = 10^{-8}$, the fixed-nuisance and S3 locally-unbiased energy scales are $5.31 \times 10^{-5}$ and 0.0699. The polynomial estimator has parameter-dependent bias, so its much smaller median does not violate the S3 bound. The Supplement gives the local response check and an independently averaged, admissible-prior van Trees illustration. These comparisons establish sensitivity to the input model, not estimator optimality.

### VI. Measurement access and copy costs

For a known pure-state ladder, let $p_n = |c_n|^2$ and $u_n = c_n c_0^*$. Calibrated pair rotations and population discrimination give the standard complementary measurements [43]

$$p^x_{n,\pm} = (p_0 + p_n)/2 \pm \mathrm{Re}u_n, p^y_{n,\pm} = (p_0 + p_n)/2 \pm \mathrm{Im}u_n. \tag{23}$$

A separate $p_0$ measurement supplies $z_n = u_n/p_0$. Allocating $S$ copies per setting at each of $N_t$ times costs $N_t(2L+1)S$ preparations for $L$ observed levels. Near preparation, the coordinate-component variance is approximately $1/(4S)$. The model assumes selective rotations, a common phase reference and independently prepared copies. Complex return-amplitude access differs from population-only access and requires phase information. Restricted-end tomography has a different acquisition input [23]. Fock-state and selective oscillator control are established resources [44, 45, 46]; their errors and overhead must be characterized for a specific implementation.

With $N_t = 61, L = 3, S = 10^{12}$, the ideal allocation costs $4.27 \times 10^{14}$ copies, or 13.5 years at a hypothetical 1 MHz repetition rate. The equal-copy information comparison in the Supplement instead uses $4.27 \times 10^{12}$ copies, about 49.4 days at that rate. Reset and pulse durations are excluded. For the same endpoint-energy target, early three-level access is about 140–510 times less informative in standard-deviation scale than the two specified later windows. Of this factor, about 14 comes from adding $z_4$, and 10–35 from later timing at fixed depth. Under exact preparation the two-site design gives the smallest bound among the specified designs for this endpoint target. Those endpoint extrapolations require S3 and do not apply to the unrestricted smooth class of Proposition 2. The Supplement also separates in-window energy targets and the two-preparation input; no experimentally attained cost follows from the local bounds.

### VII. Additional preparations and onset aliases

Proposition 4 (second-order site-preparation access). Prepare $|k\rangle$ and measure an adjacent ratio $w_{j\leftarrow k} = c_j/c_k$, $|j-k| = 1$. At preparation,

$$w' = -\frac{i g_{\min(j,k)}}{\hbar}, \frac{w''}{w'} = \frac{\dot{g}_{\min(j,k)}}{g_{\min(j,k)}} - \frac{i(E_j - E_k)}{\hbar}. \tag{24}$$

A coherent preparation admixture requires calibration or a preparation-aware fit with a free constant. Each site covers its incident edges. For arbitrary independent rates on $M$ edges, $\lceil M/2 \rceil$ site preparations are necessary and sufficient at second order: the selected vertices must cover the path. Arbitrary superposition inputs have different requirements [46]. For the first three edges, sites 0,2 suffice. For the oscillator these are the onset form of Fock-resolved two-photon Rabi/Ramsey relations under the specified pair readout, with a lower derivative order bought by preparation access.

Proposition 5 (dissipative onset aliases). Let $H(t)$ preserve photon-number parity, with a real Jacobi even block and positive relevant onset couplings, whose edge $k$ joins $|2k\rangle$ and $|2k+2\rangle$. Define $\chi_k = \dot{g}_k/g_k$ and $\mathcal{D}[L]\rho = L\rho L^\dagger - \{L^\dagger L, \rho\}/2$, and assume $\dot{\rho} = -i[H,\rho]/\hbar + \kappa_-\mathcal{D}[a]\rho + \kappa_+\mathcal{D}[a^\dagger]\rho + \gamma\mathcal{D}[\hat{n}]\rho$. Here $\gamma\mathcal{D}[\hat{n}]$ damps a coherence of number separation $\Delta n$ at rate $\gamma(\Delta n)^2/2$. For preparation $|2k\rangle$ and $w_k = \rho_{2k+2,2k}/\rho_{2k,2k}$,

$$\operatorname{Re}\frac{w_k''}{w_k'} = \chi_k - \kappa_- - \kappa_+ - 2\gamma. \tag{25}$$

All quantities are evaluated at preparation. The odd block can be arbitrary subject to parity preservation. For the banded pair-driven oscillator $H = D(\hat{n},t) + \hbar G(t)(a^2 + a^{\dagger 2})$, vacuum ratios $\rho_{2n,0}/\rho_{00}$ give

$$\hat{\chi}_{n-1}^{\text{onset}} = \chi_{n-1} - \kappa_- - \kappa_+ - 2n\gamma, \tag{26}$$

where "onset" denotes the inference from the first two nonzero endpoint coefficients. The first identity allows differences of independent local rates when preparation families share the same generator and dissipative rates. For a pure pair drive all $\chi_k = \dot{G}/G$ coincide, so a nonzero rate difference tests departure from that model. Two-photon loss $\kappa_2\mathcal{D}[a^2]$ adds shifts $-\kappa_2(4k+1)$ and $-\kappa_2(2n-1)$ to the site and vacuum inferred rates, respectively; the site-4 minus site-0 difference retains $-8\kappa_2$. Nonlinear number dephasing also leaves edge-dependent shifts. The Supplement gives the proofs and cutoff conditions. Preparation families require matched drive onset and readout calibration. A zero-onset ramp falls outside the positive-onset identities above.

## VIII. Proofs

### A. Endpoint identities

Set $t_0 = 0$ and use the projective scalar gauge $E_0(0) = 0$. Dividing the Schrödinger amplitudes by $c_0$ gives, for $n \geq 1$,

$$i\hbar\dot{z}_n = (E_n - E_0)z_n + g_{n-1}z_{n-1} + g_n z_{n+1} - g_0 z_1 z_n. \tag{27}$$

Nearest-neighbour topology implies $z_n(t) = a_n t^n + b_n t^{n+1} + O(t^{n+2})$. Since $z_1 = O(t)$, the nonlinear projective term $z_1 z_n$ starts at $O(t^{n+1})$ and does not enter the coefficients of $t^{n-1}$ or $t^n$ used below. The $t^{n-1}$ coefficient gives

$$i\hbar n a_n = g_{n-1} a_{n-1}. \tag{28}$$

With $a_0 = 1$ and $b_0 = 0$, recursion yields $A_n = n!\, a_n = (-i/\hbar)^n \prod_{j=0}^{n-1} g_j$. At order $t^n$:

$$i\hbar(n+1)b_n = E_n a_n + g_{n-1} b_{n-1} + \dot{g}_{n-1} a_{n-1}. \tag{29}$$

Because $R_n = (n+1)b_n/a_n$, equation (28) reduces (29) to

$$R_n = R_{n-1} + n\frac{\dot{g}_{n-1}}{g_{n-1}} - \frac{i}{\hbar}E_n. \tag{30}$$

Restoring the projective scalar gauge replaces $E_n$ by $E_n - E_0$. For a real chain, $\dot{g}/g$ is real, so the real and imaginary parts give (7)-(9). Derivatives of the diagonal energies enter only at higher projective order and are not needed for instantaneous layer reconstruction. The smoothness assumption in Theorem 1 ensures the classical derivative jet and the stated Taylor remainders.

**B. Necessity of the derivative order**

Repeated differentiation of (2) gives $\psi^{(k+1)}(t_0) = -(i/\hbar)\sum_{j=0}^{k}\binom{k}{j} H^{(j)}(t_0)\psi^{(k-j)}(t_0)$. A term sensitive to a constant shift of $E_M$ must reach level $M$ and then apply that diagonal entry. The shortest such term has $M$ nearest-neighbour steps and one diagonal step, so no derivative through order $M$ changes. At order $M+1$, the affected component is precisely $(-i/\hbar)^{M+1}\delta \prod_{j=0}^{M-1} g_j\,(t_0)$. Division by $(M+1)!$ gives (10). The denominator $c_0$ cannot alter this leading coefficient. An affected return path to level zero requires $2M+1$ steps. Thus the entire projective jet through order $M$ is identical while the deepest energy differs. This proves necessity in the unrestricted class. A change in $\dot{g}_{M-1}(t_0)$ likewise first enters at order $M+1$.

## IX. Discussion

The results address three distinct aspects of the inverse problem: uniqueness from an exact endpoint jet, nonuniqueness under restricted observation, and conditioning under finite-time noisy sampling. Theorem 1 gives a constructive, depth-by-depth inversion and the sharp derivative order for complete instantaneous layers. Theorem 2 shows that an arbitrarily long Taylor jet of a restricted coordinate prefix does not determine hidden layers when the coefficient time laws are unrestricted and smooth. Proposition 2 gives a different obstruction: even full-state observation on a window separated from preparation can be compatible with distinct endpoint coupling scales and rates. Consequently, increasing temporal resolution or observation duration does not in itself replace structural assumptions or access to missing data. Generic-time reconstruction and additional site preparations offer different measurement operators with lower derivative requirements; they should be chosen according to the accessible observables and parameter targets.

These conclusions depend on the following assumptions. Exact endpoint inversion uses a fixed real-Jacobi basis, positive relevant couplings, a calibrated physical time coordinate, and phase-resolved projective data. In complex chains, a changing coupling phase is indistinguishable from a diagonal-energy contribution without a phase-gauge specification. Smooth but unrestricted time dependence also differs from analytic, polynomial or derivative-bounded coefficient classes. Finite-window estimators require their own approximation, preparation and noise controls; the reported numerical results are synthetic demonstrations within specified quadratic models, not an experimental validation. Applications include Hamiltonian identification in tight-binding models and the even-Fock ladders of parametrically driven bosonic systems, subject to the stated measurement-access requirements.

For the fixed real-Jacobi class, the main contribution is a phase-resolved endpoint reconstruction with a sharp derivative-order threshold, together with exact nonuniqueness results under incomplete observation. The construction specializes state-derivative Hamiltonian reconstruction to singular endpoint data. Projective geometry motivates the observation model; the identifiability results are derived from Schrödinger evolution.

## X. Conclusion

We establish a sharp reconstruction threshold for time-dependent real Jacobi Hamiltonians from endpoint projective trajectory jets: order $M+1$ identifies the first $M$ complete instantaneous layers, whereas unrestricted smooth time laws prevent deeper reconstruction from all-order jets of a restricted observed prefix. A separate exact observation-gap construction demonstrates endpoint ambiguities even with complete later-time state access. Generic-time population-assisted data and multiple site preparations reduce derivative requirements under their respective nondegeneracy and access assumptions. These uniqueness and nonuniqueness results, combined with local statistical bounds and finite-window simulations, provide a structured basis for designing and interpreting inverse Hamiltonian measurements.

**Acknowledgments**

The author used generative AI tools for language editing, code assistance and critical review, and takes responsibility for the content, accuracy and integrity of the work.

**Data and code availability**

All code and data are available at https://github.com/alexandermtishin-art/Local-projective-reconstruction-of-driven-bosonic-Hamiltonians . The archive contains scripts, fixed seeds, environment records and stored outputs for the stated controls. Read-only integrity checks and recomputation commands are documented separately. No experimental trajectory is reconstructed.